\documentclass{article}

\usepackage{arxiv}

\usepackage[utf8]{inputenc} 
\usepackage[T1]{fontenc}    
\usepackage[hidelinks]{hyperref}       
\usepackage{url}            
\usepackage{booktabs}       
\usepackage{amsfonts}       
\usepackage{amssymb,amsmath}
\usepackage{nicefrac}       
\usepackage{microtype}      
\usepackage{lipsum}
\usepackage{graphicx}
\usepackage{xcolor}

\usepackage{longtable,booktabs}
\usepackage{graphicx,grffile}

\usepackage{titling}
\newcommand{\spA}{\texttt{A}}
\newcommand{\spB}{\texttt{B}}
\newcommand{\spC}{\texttt{C}}
\newcommand{\spD}{\texttt{D}}
\newcommand{\spE}{\texttt{E}}

\usepackage{wrapfig}
\usepackage{booktabs}
\usepackage[numbers]{natbib}
\title{Novel Symmetry-preserving Neural Network Model for Phylogenetic Inference}

\author{
Xudong Tang\thanks{Equal contribution; authors in alphabetical order} \\
  Wisconsin Institute for Discovery\\
  University of Wisconsin-Madison\\
  Madison, WI 53706 \\
  \and
    Leonardo Zepeda-Nu\~{n}ez$^*$\thanks{Now at Google Research} \\
  Department of Mathematics\\
  University of Wisconsin-Madison\\
  Madison, WI 53706  
  \and
  Shengwen Yang\\
  Department of Statistics\\
  University of Wisconsin-Madison\\
  Madison, WI 53706 \\
  \and
    Zelin Zhao\\
  Department of Mathematics\\
  University of Wisconsin-Madison\\
  Madison, WI 53706 \\
  \and
 Claudia Sol\'{i}s-Lemus\thanks{Corresponding author: solislemus@wisc.edu} \\
  Wisconsin Institute for Discovery\\
  University of Wisconsin-Madison\\
  Madison, WI 53706 \\
}

\date{}

\begin{document}
\maketitle

\begin{abstract}
Scientists world-wide are putting together massive efforts to understand how the biodiversity that
we see on Earth evolved from single-cell organisms at the origin of life and this diversification process
is represented through the Tree of Life. Low sampling rates and high heterogeneity in the rate of evolution across sites and lineages produce a phenomenon denoted ``long branch attraction" (LBA) in which long non-sister lineages are estimated to be sisters regardless of their true evolutionary relationship. LBA has been a pervasive problem in phylogenetic inference affecting different types of methodologies from distance-based to likelihood-based. Here, we present a novel neural network model that outperforms standard phylogenetic methods and other neural network implementations under LBA settings. Furthermore, unlike existing neural network models, our model naturally accounts for the tree isomorphisms via permutation invariant functions which ultimately result in lower memory and allows the seamless extension to larger trees.
\end{abstract}


\section{Introduction}
The Tree of Life is the representation of the evolutionary process that originated the world-wide diversity of life from single-cell organisms. Existing approaches to reconstruct the Tree of Life involve the collection of genomic data from living organisms, and then, the use of statistical techniques to estimate the tree which is the best fit to the observed genomic data under a specified model of evolution.
While the cost of sequencing collected organisms continues to decrease, many clades in the tree still do not have a high sampling proportion due to limited or inaccessible species ranges or poor allocation of resources. Furthermore, in addition to the 1.4 million estimated species, it is estimated that there are from 10 to 100 million undiscovered and undescribed species \cite{Groomberg1992}.

Low sampling of species can cause long branches in the tree which could be incorrectly grouped in a phenomenon denoted ``long branch attraction" (LBA) \cite{hendy1989framework, felsenstein1978cases, bergsten2005review, anderson2004should}
where long branches in the tree are estimated to be sisters regardless of their true evolutionary relationship. More formally, LBA arises when the probability that close relatives share character states due to common ancestry is exceeded by the probability that more distantly related taxa share states due to convergent evolution \cite{hendy1989framework}. It has been studied that increasing the sampling frequency can potentially divide those long branches and thus, overcome the LBA problem \cite{gauthier1988amniote, wiens2005can}, yet it is not always possible to increase sampling of some clades in the Tree of Life.

Aside from low sampling, LBA also arises when there are inequalities in the rate of evolution among branches \cite{felsenstein1978cases}. In this scenario, commonly referred to as ``Felsenstein zone" \cite{felsenstein2004inferring, huelsenbeck1993success}, an incorrect tree that groups taxa belonging to high-rate lineages will be more likely to be estimated. This implies that LBA would appear in rapidly evolving lineages which produce long branches leading to some taxa which are then artificially estimated as sister taxa by most traditional phylogenetic methods. These traditional methods tend to ignore heterogeneities such as within-site rate variation (heterotachy) \cite{philippe2005heterotachy} or rate variation across lineages.
Site-heterogeneous models \cite{lartillot2004bayesian} that allow different rates of mutation for different sites have successfully overcome LBA for some datasets \cite{lartillot2007suppression}. Yet the complexity to infer site-heterogeneous models has hampered its use for large datasets with many sites and many taxa.

The first implementations of neural network models in phylogenetic inference \cite{Suvorov2020-pl, Zou2020-ta} have found that neural networks are able to accurately estimate phylogenetic trees with long branches and thus, overcome the LBA phenomenon. 
These implementations share three similarities: 1) they all focus on neural network models (mainly CNNs); 2) they can only estimate 4-taxon trees, and 3) they have been proven to outperform standard phylogenetic inference methods like maximum likelihood, Bayesian inference and maximum parsimony under a variety of heterogeneous simulated scenarios, especially those related to the anomaly zone (LBA) \cite{Suvorov2020-pl, Zou2020-ta}, Felsenstein trees, Farris trees \cite{leuchtenberger_distinguishing_2020} or scenarios with highly gapped alignments \cite{Suvorov2020-pl}.

The main shared weakness of these phylogenetic neural network models is their limitation to 4-taxon datasets. While Zou \textit{et al} \cite{Zou2020-ta} was able to analyze larger real datasets in combination with Quartet Puzzling \cite{strimmer1996quartet}, extending the neural network model to $n$ taxa would allow a direct estimation of a $n$-taxon tree as opposed to breaking up an $n$-taxon dataset into all combinations of 4 taxa to later merge the resulting quartets.
One of the challenges to extend these neural network models to more taxa is the symmetric nature of the tree structure. 
For example, in the quartet \texttt{((A,B),(C,D))}, the sequences \spA\ and \spB\ (and the sequences \spC\ and \spD) are interchangeable, as are the clades \texttt{(A,B)} and \texttt{(C,D)} with \texttt{((C,D),(A,B))} or \texttt{((C,D),(B,A))} representing the same tree structure.
Zou \textit{et al} \cite{Zou2020-ta} account for all the symmetries by artificially creating all 24 permutations of the sequences to guarantee that their neural network model is invariant to the symmetries.
This process, however, is not scalable for larger trees both in terms of computing time and memory allocation for all the artificial permutations.
More recent implementations of neural networks in phylogenetics can estimate trees with up to 15 taxa \cite{Smith2022}, yet this model relies on comparing simulated data with real data, rather than classifying the best tree. Other deep learning approaches, such as the \texttt{Phyloformer} \cite{Nesterenko2022} could infer much larger trees. However, this approach apply deep learning on constructing the pairwise distance matrix, thus it is not a direct estimation of the tree topologies by the model itself.

Here, we exploit the inherent symmetries of the tree objects with the first symmetry-preserving neural network model which provides a classification framework that can be extended to more than four taxa. 
We show that our NN model not only outperforms standard phylogenetic inference methods like neighbor joining, maximum likelihood and Bayesian inference, it also outperforms existing neural network models \cite{Zou2020-ta} all while reducing the memory needs and training time by not requiring the creation of artificial permutations of the data. Our work opens the door to neural network models capable of directly classifying $n$-taxon trees through the invariant transformations that are built into our model.


\section{Description of symmetry-preserving neural network model for four taxa}
\label{sec:NN_model}

The particular structure of our proposed method leverages two main features of the problem: the positional information within the sequence and the equivariance of the tree-topology with respect to the order of the sequences. These two main features are then embedded within the architecture of the network.
In what follows we will describe in general terms the architecture and how is able to preserve the two properties mentioned above. For the sake of clarity, we will present them in a modular fashion, similarly as it is implemented in the code. 

\begin{figure}[h]
\centering
\includegraphics[width=8cm]{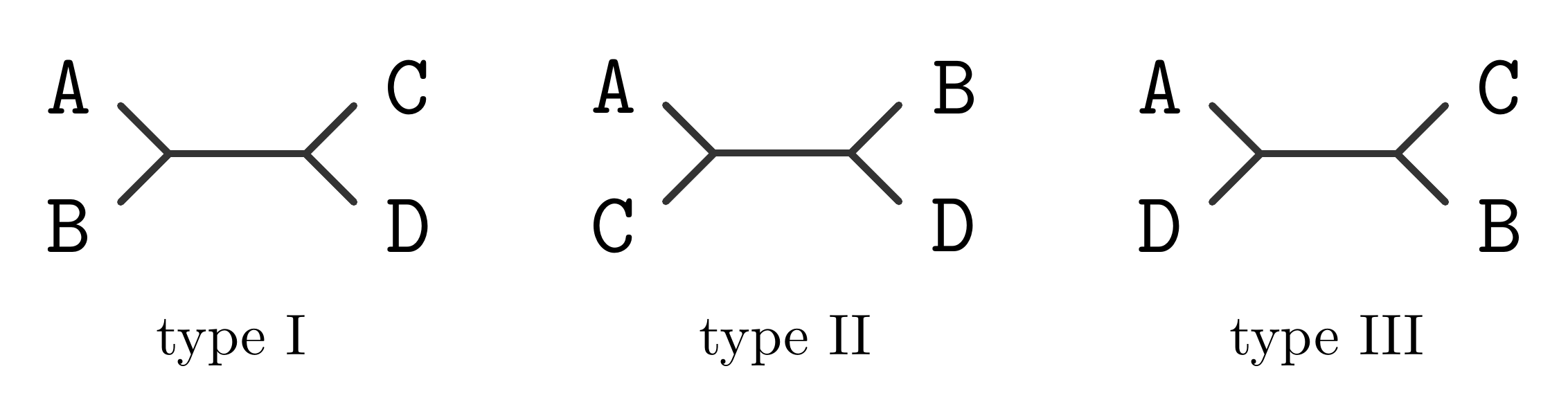} \hspace{0.15cm}
\includegraphics[width=8cm]{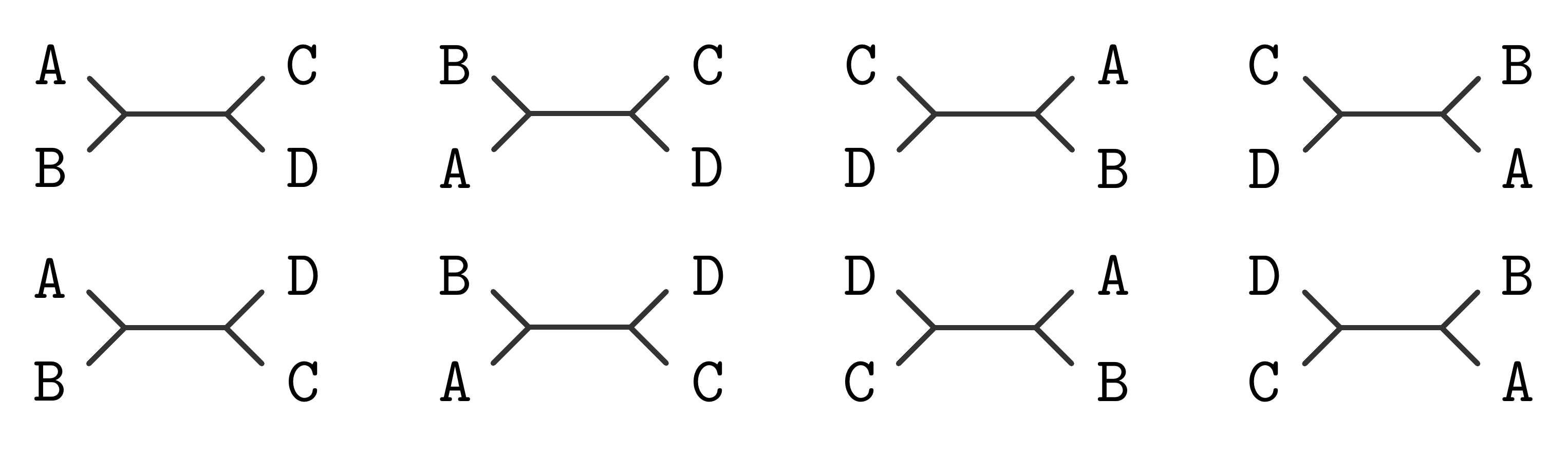}
\caption{Top: Possible unrooted trees for 4 species (quartets). Bottom: All the possible permutations on the order of the leaves that are isomorphic to a tree of type I.}
\label{fig:trees}
\end{figure}

\noindent \textbf{Symmetry-preserving neural network.}
Even though the number of possible trees increases super-exponentially with the number of species, there exist several symmetries that one can leverage. Suppose that we have four different species, denoted \spA, \spB, \spC, and \spD, and for each of these species we have data in the form of sequences $s_{\spA}$, $s_{\spB}$, $s_{\spC}$, and $s_{\spD}$. The task at hand is to estimate the most likely tree topology that explains the data fed as a matrix of the form $[s_{\spA}, s_{\spB},s_{\spC}, s_{\spD}]$. In this case of four taxa, we only have \emph{three} possible unrooted tree topologies which are depicted in Fig.~\ref{fig:trees} (top). We then seek a function $p([s_{\spA}, s_{\spB},s_{\spC}, s_{\spD}]) \rightarrow [p_I, p_{II}, p_{III}]$, where $p_I$ is the probability that the data is explained by the tree of type I (and similarly for $p_{II}$ and $p_{III}$). Even though we can obtain $24$ different trees with four leaves by permutation of the leaf labels, many of these are 
equivalent in the sense that they share the same labelled topology, as shown in Fig.~\ref{fig:trees} (bottom).

This simple example already sheds light on the properties we will exploit
to reduce the search space. The order in which the input is fed is completely arbitrary, and it is subject to several invariances. For example, if the tree to be estimated is of type I, then exchanging the position of $\spA$ and $\spB$ (or $\spC$ and $\spD$) leads to the same topology (Fig.~\ref{fig:trees} top). Thus, if we were to feed our algorithm the data ordered as $[s_{\spB}, s_{\spA},s_{\spC}, s_{\spD}]$, or alternatively as $[s_{\spA}, s_{\spB},s_{\spD}, s_{\spC}]$, we would expect the score $p_I$ to be the same. In addition, if we exchange each clade, i.e., we swap $\spA$ and $\spB$ by $\spC$ and $\spD$, then the 
topology remains the same, and so should the score $p_I$.
In addition, if one tree is predicted to be certain type, and then we permute the order in which the sequences are feed, it may change the type that was predicted originally. In fact, there is a correspondence between how the labels should be modified under any permutation of the sequences.

Creating a network that is invariant to these
transformations will reduce the number of data points necessary for training which in turn would reduce the computational cost. 
More formally, we will take advantage of the newly developed theory of invariant networks~\cite{ZaheerKotturRavanbakhshEtAl2017}, which states that any permutation invariant function $f:\mathbb{R}^n \rightarrow \mathbb{R}$ accepts a representation of the form 
\begin{equation*}
    f(x_1, x_2, x_3, ..., x_n) = \Phi \left (\sum_{i = 1}^{n} \phi(x_i) \right) \,\,\, 
\end{equation*}
where
\begin{align*}
\begin{split}
    \phi&: \mathbb{R} \rightarrow \mathbb{R}^m \,\, \\
    \Phi&:\mathbb{R}^m \rightarrow \mathbb{R}^m
\end{split}
\end{align*}
where $m$ is the embedding dimension, which is case dependent. In our study, we found out that $m=80$ offers the best balance between performance and scalability for our input dimension. We will exploit this theory of invariant networks and extend it to the case of phylogenetic trees.

For example, suppose that $\phi: \mathbb{D}^{L} \rightarrow \mathbb{R}^{L \times m} $, where $\mathbb{D}$ is the dictionary for the sequences (e.g., the set of nucleotides or amino acids), and $L$ is the sequence length. Suppose that $\Phi: \mathbb{R}^{L \times m} \rightarrow \mathbb{R}^{L \times m}$, then we have that $\Phi(\phi(s_{\spA}) + \phi(s_{\spB}))$ will be invariant if we permute $\spA$ and $\spB$, and $\Phi(\phi(s_{\spC}) + \phi(s_{\spD}))$ will be invariant if we permute $\spC$ and $\spD$. 

Then, we define the descriptor
\begin{equation*}
\begin{split}
     \mathcal{D}_{I}([s_{\spA}, s_{\spB},s_{\spC}, s_{\spD}])  &= \Phi(\phi(s_{\spA}) + \phi(s_{\spB}))\\&+  \Phi(\phi(s_{\spC}) + \phi(s_{\spD}))
\end{split}
\end{equation*}
which is clearly invariant to the transformations which are invariant for the tree of type I. 

Thus, if we have a third function $\Psi: \mathbb{R}^{L \times m} \rightarrow [0,1]$, and we can write
\begin{align*}
\begin{split}
p_{I}([s_{\spB}, s_{\spA},s_{\spC}, s_{\spD}]) &=  \Psi(\Phi(\phi(s_{\spA}) + \phi(s_{\spB}))+ \Phi(\phi(s_{\spC}) + \phi(s_{\spD})))\\
&= \Psi(\mathcal{D}_{I}([s_{\spB}, s_{\spA},s_{\spC}, s_{\spD}]) ) 
\end{split}
\end{align*}
which is a score function that satisfies all the symmetries required. 
We use the same strategy to define the scores for trees of type II and III, resulting in 
\begin{align*}
p_{II}([s_{\spB}, s_{\spA},s_{\spC}, s_{\spD}]) &= \Psi(\Phi(\phi(s_{\spA}) + \phi(s_{\spC}))\\&+  \Phi(\phi(s_{\spB}) + \phi(s_{\spD}))), \\
p_{III}([s_{\spB}, s_{\spA},s_{\spC}, s_{\spD}]) &=  \Psi(\Phi(\phi(s_{\spA}) + \phi(s_{\spD}))\\&+  \Phi(\phi(s_{\spC}) + \phi(s_{\spB}))).
\end{align*}

The scores can be written in a more compact fashion by aggregating them into a vector:
\begin{equation} \label{eq:scores}
    p( [s_{\spB}, s_{\spA},s_{\spC}, s_{\spD}]) = \left [ \begin{array}{c}
         p_{I}  \\
         p_{II} \\
         p_{III}
    \end{array} \right] = 
        \left [ \begin{array}{r}
         \Psi(\mathcal{D}_{I} ([s_{\spA}, s_{\spB},s_{\spC}, s_{\spD}]) ) \\
         \Psi(\mathcal{D}_{II} ([s_{\spA}, s_{\spB},s_{\spC}, s_{\spD}]))  \\
         \Psi(\mathcal{D}_{III} ([s_{\spA}, s_{\spB},s_{\spC}, s_{\spD}]))
    \end{array} \right]
\end{equation}
where the descriptors are given by
\begin{equation} \label{eq:descriptors}
  \left [ \begin{array}{c}
         \mathcal{D}_{I}  \\
         \mathcal{D}_{II} \\
         \mathcal{D}_{III}
    \end{array} \right] = 
        \left [ \begin{array}{r}
         \Phi(\phi(s_{\spA}) + \phi(s_{\spB}))+  \Phi(\phi(s_{\spC}) + \phi(s_{\spD})) \\
         \Phi(\phi(s_{\spA}) + \phi(s_{\spC}))+  \Phi(\phi(s_{\spB}) + \phi(s_{\spD}))  \\
         \Phi(\phi(s_{\spA}) + \phi(s_{\spD}))+  \Phi(\phi(s_{\spC}) + \phi(s_{\spB}))
    \end{array} \right].
\end{equation}

This scoring function satisfies all the symmetries, and it only requires us to find three functions, which will be parametrized using NNs. This technique can be generalized to larger numbers of species seamlessly.

In what follows we provide extra details on how the descriptor $\mathcal{D}$ and the scoring function $\Psi$ are implemented. 

\begin{figure*}[!h]
\centering
\includegraphics[scale=0.15]{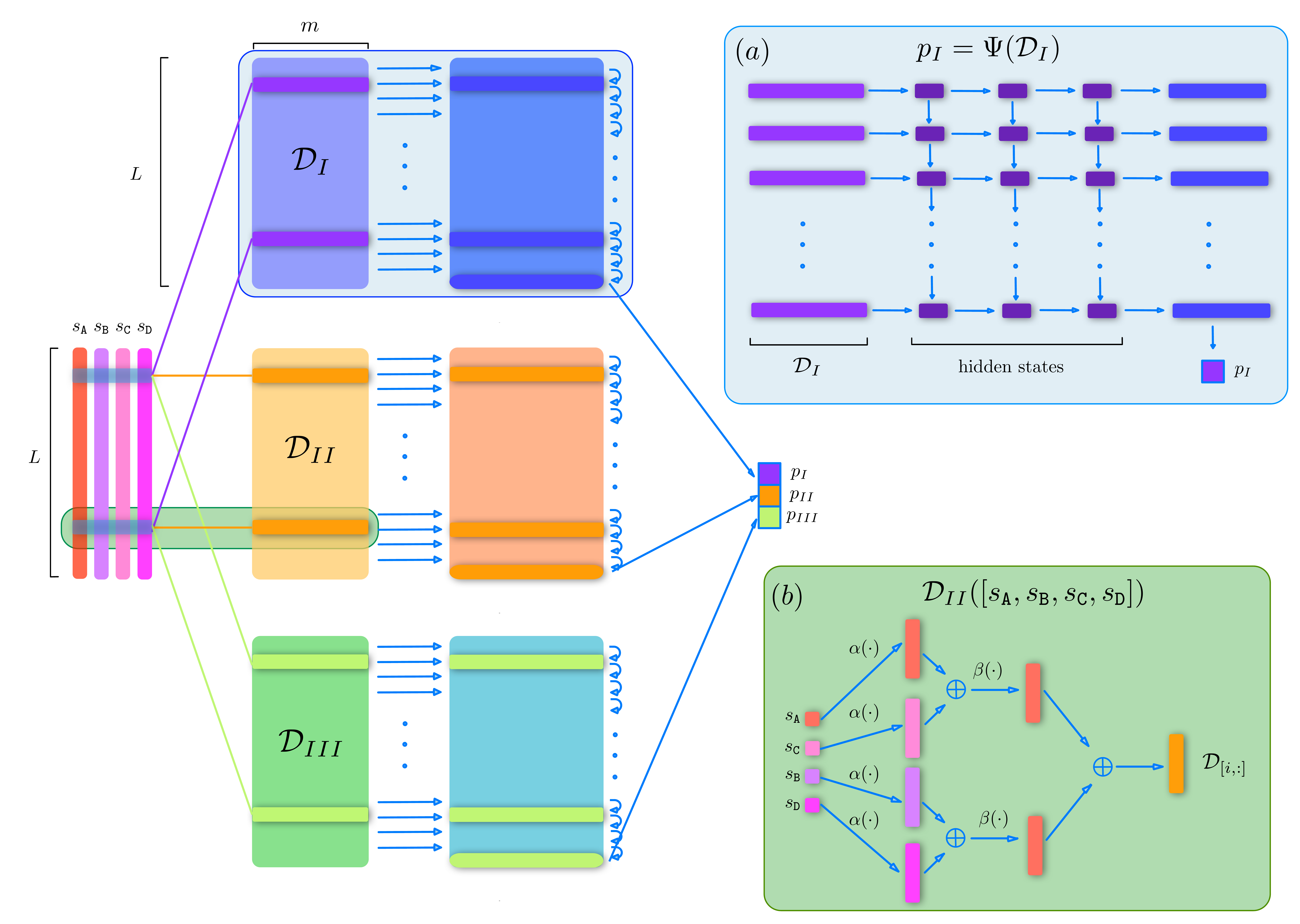}
\caption{Diagram of the network. We start with sequences $[s_{\spA}, s_{\spB},s_{\spC}, s_{\spD}]$ in the left, then we apply the same functions $\phi$ and $\Phi$, but in a different order to obtain the descriptors $\mathcal{D}_I$, $\mathcal{D}_{II}$, and $\mathcal{D}_{III}$, which are described in (b). Then each descriptor is fed to the function $\Psi$ (shown in (a)) which outputs the score for each particular tree topology.
(a) Diagram of $\Psi$ which is described by a LSTM followed by a dense network.
(b) Diagram of the computation of a slice of $\mathcal{D}_{II}$, relying on the functions $\alpha$, which is an embedding layer, and $\beta$, which is a shallow dense ResNet. 
}
\label{fig:network}
\end{figure*}

\vspace{0.2cm}
\noindent \textbf{Symmetry-preserving descriptors.}
We now provide the exact form of the descriptors for each tree topology. The main motivation of each descriptor is to respect both the symmetries and the ordering imposed by the sequences. The descriptor will apply the same transformation to each site but across the four sequences independently of the rest of the sites. In a nutshell, the $i$-th slice of the descriptor will only depend on the $i$-th site of the input sequences.

In particular, the descriptor take an input of the form
\begin{equation*}
    \mathcal{D}_{I}: \mathbb{D}^{L \times 4} \rightarrow \mathbb{R}^{L \times m}
\end{equation*}
by applying the same transformation to each of the slices of the input $[s_{\spA}, s_{\spB},s_{\spC}, s_{\spD}]$ independently. 

By the sake of simplicity, we will denote by a generic $s \in \mathbb{D}^{L}$ one of the four sequences mentioned before and  $s_i \in \mathbb{D}$ its $i$-th component (or site). 

As shown in Fig.~\ref{fig:network} (b), the first operation is an embedding layer, which will use the map $\alpha:\mathbb{D} \rightarrow \mathbb{R}^m$, and it will be applied to each site of the sequence, i.e., the function $\phi$ mentioned above
\begin{equation*}
    \left (\phi (s)\right)_{[i,:]} = \alpha(s_i)
\end{equation*}
where we have used Matlab notation (i.e., if $A \in \mathbb{R}^{m\times m}$ then $A_{[i,:]}$ corresponds to the $i$-th row of $A$). 

Then, we can do the same for $\Phi$, which will be written as the application of a function $\beta:\mathbb{R}^m \rightarrow \mathbb{R}^m$ to each slice of the output of $\phi$, i.e., 
\begin{equation*}
\begin{split}
    \left (\Phi (\phi(s) + \phi(s'))\right)_{[i,:]} &=  \beta\left(\left(\phi(s)_{[i,:]} + \phi(s')\right)_{[i,:]} \right)\\& = \beta(\phi(s)_{[i,:]} + \phi(s')_{[i,:]}) \\&= \beta(\alpha(s_{i}) + \alpha(s'_{i})).
\end{split}
\end{equation*}

Then, we can write a generic descriptor by slice as 
\begin{equation*}
    \mathcal{D}_{[i,:]} = \beta(\alpha(s_{i}) + \alpha(s'_{i})) + \beta(\alpha(\tilde{s}_{i}) + \alpha(\hat{s}_{i})),
\end{equation*}
where $s, s', \tilde{s},$ and $\hat{s}$ are a certain ordering of the generic sequences $s_{\spA}, s_{\spB},s_{\spC},$ and $s_{\spD}$, operation depicted in Fig.~\ref{fig:network} (b). 

We add the correct ordering of the sequences to take place in the expression above. This produces three different sequences of descriptors, each one of dimension $L \times m$. In this case, $\alpha$ is implemented by an embedding layer. This choice allows us to avoid the hot-encoding, which allows in turn to feed the data to the network more efficiently. On the other hand, $\beta$ implemented by dense ResNet block~\cite{HeZhangRenEtAl2016}. One could add a deeper network for the $\alpha$ and $\beta$ functions. However, empirically, we have found that gains seem to be marginal.

For simplicity $\phi$ and $\Phi$ are implemented with one-dimensional convolutions with a kernel size of one, instead a dense network. This produces the same effect as mentioned above, but allows us to avoid any unnecessary reshape and efficiently batch the computations. 

We point out that the descriptors maintain the {\it order} in the sequence, which will be then leveraged by the recurrent network in what follows. 

\vspace{0.2cm}
\noindent \textbf{Recurrent network.} 
Given the ordered nature of the data and the descriptor network that was built to preserve such order, the scoring function $\Psi$ is implemented using a standard recurrent LSTM network~\cite{LSTM:1997} using the PyTorch built-in models to process the sequence of each descriptor to another sequence and then use the last element of the new sequence to compute the score as depicted in Fig.~\ref{fig:network} (a). 

As shown in Fig.~\ref{fig:network} (a), the LSTM network takes each of the slices of the descriptor at each site, and the hidden states of the precedent slice, and outputs a new descriptor that contains the information of the current slice and the information carried over by the hidden states of the precedent slices, while preserving some of the information in a hidden state for the next slice. 
This is repeated throughout the sequence, resulting in a new sequence of descriptors in which at each site, unlike the input, has information about all the sites {\it before it} and thus accounting for correlation among sites in one direction. Then we only consider the last row (or slice) of the output of the LSTM, which is fed to a feed-forward neural network, and ultimately outputs one single number: the score of the network to be of certain tree type. 

The same process is repeated for the descriptors of each type of tree in Equation \ref{eq:descriptors} and their outputs are then concatenated, thus obtaining the scores following Equation \ref{eq:scores}, which is shown in Fig.~\ref{fig:network}.

\vspace{0.2cm}
\noindent \textbf{Loss function and optimizer.} 
We used a typical multi class cross entropy loss, and the optimizer used was Adam~\cite{adam} with the default parameters in PyTorch~\cite{Pytorch}. We used an exponential scheduler for the learning rate.

\subsection{Extension to five taxa}
\begin{figure}[h]
\centering
\includegraphics[scale=0.3]{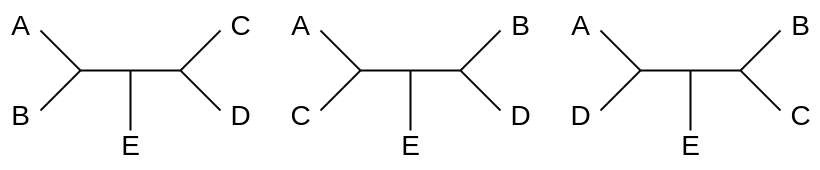}
\caption{Three possible 5-taxon unrooted tree topologies out of the 15 possibilities. Note that we can still exploit the symmetry characteristic of these topologies.}
\label{fig:quintet-tree}
\end{figure}

Let $\spA, \spB, \spC, \spD$, and $\spE$ be the five taxa with sequences 
$s_{\spA}$, $s_{\spB}$, $s_{\spC}$, $s_{\spD}$, and $s_{\spE}$.
While for four taxa there are 3 possible unrooted tree topologies, there are now 15 possibilities for the case of five taxa (Fig.~\ref{fig:quintet-tree}). We can group these trees by the one species in the middle (in Fig.~\ref{fig:quintet-tree}, it is taxon $\spE$) having three different topologies. 

We can still exploit the
symmetric characteristic in quartet trees when working with quintet trees. For example, in Fig.~\ref{fig:quintet-tree}, species $\spE$ is in the middle. We could simply ignore species $\spE$ and represent the two binary tree tips on both sides using $\Phi$ and $\phi$ functions as we did in the quartet tree cases. Essentially, we would create 15 descriptors on 15 quartet trees coming from the original 15 quintet trees excluding the middle species as in Equation \ref{eq:descriptors_five}:
\begin{equation} \label{eq:descriptors_five}
  \left [ \begin{array}{c}
         \mathcal{D}_{I}  \\
         \mathcal{D}_{II} \\
         \mathcal{D}_{III} \\
         \mathcal{D}_{IV} \\
         \text{...} \\
         \mathcal{D}_{XV}
    \end{array} \right] = 
        \left [ \begin{array}{r}
         \Phi(\phi(s_{\spA}) + \phi(s_{\spB}))+  \Phi(\phi(s_{\spC}) + \phi(s_{\spD})) \\
         \Phi(\phi(s_{\spA}) + \phi(s_{\spC}))+  \Phi(\phi(s_{\spB}) + \phi(s_{\spD}))  \\
         \Phi(\phi(s_{\spA}) + \phi(s_{\spD}))+  \Phi(\phi(s_{\spC}) + \phi(s_{\spB})) \\
         \Phi(\phi(s_{\spA}) + \phi(s_{\spB}))+  \Phi(\phi(s_{\spC}) + \phi(s_{\spE})) \\
         \text{...} \\
         \Phi(\phi(s_{\spB}) + \phi(s_{\spC}))+  \Phi(\phi(s_{\spD}) + \phi(s_{\spE}))
    \end{array} \right].
\end{equation}

The model components and structures are exactly the same as in Fig.~\ref{fig:network} except now we have 15 Descriptors and 15 $p$ values.

\subsection{Open-source implementation}
Our NN model was implemented using Python leveraging Pytorch. We implement two version of the network, dubbed in what follows, LSTM and OptLSTM. The first one showcases the proposed network, whereas the second is an optimized version of the pipeline. 
All the scripts for the model implementation can be found in the GitHub repository: 
\url{https://github.com/crsl4/nn-phylogenetics}.

\section{Simulations}
\label{sims}

\subsection{Simulation strategy on four taxa}

We simulate protein sequences on known four-taxon trees. Then, we use the sequences as input data and the tree topology as labels to train the neural network model described in Section \ref{sec:NN_model}.

We follow the simulation procedure in \cite{Zou2020-ta}.
Zou \textit{et al} simulate protein sequences under the CTMC model and quartet gene trees, and then train a neural network model to classify sequences based on the quartet gene tree they came from.
A tree out of the three possible unrooted quartets was selected at random 
and we simulate branch lengths under long-branch attraction cases: trees with two long branches, each sister to a short branch and separated by a short internal branch \cite{Felsenstein1978-bq}). 
In this setting, we simulate the short external branches $b=0.1,0.5,1.0$ and then set the two long branches as $a=\kappa_1 b$ for $\kappa_1=2,10,40$ and the internal branch as $c= \kappa_2 b$ for $\kappa_2=0.01,0.1,1.0$ (see Fig.~\ref{fig:acc} top). These settings mimic the long-branch attraction simulations in \cite{Zou2020-ta}.

The quartet was rooted in the internal branch always. A sequence of length $L=1550$ (randomly chosen length in the window $(100,3000)$ used in \cite{Zou2020-ta}) is simulated following a continuous-time Markov model using PAML \cite{Yang1998-qw, Yang2007-dg}. We use the amino acid substitution empirical model of Dayhoff \cite{Dayhoff1978-fe} which was randomly chosen among the empirical models available in PAML.
For every site, a relative evolutionary rate $r$ was factored into the branch lengths. This rate was sampled from a uniform distribution in $(0.05,1.0)$ as in \cite{Zou2020-ta} and used in all the sites (that is, no site-specific rates were used).

The raw input data consisted in four amino acid sequences of length $L$ which are fed through an embedding layer.
Unlike \cite{Zou2020-ta}, we do not have to permute the data to account for the symmetries. Our neural network model incorporates the symmetries directly (as described in the model description).

For these experiments we fixed the embedding dimension $m = 80$, and an hidden dimension of the LSTM network of $20$ with a drop out probability of $0.2$, a hidden layer between the LSTM layer and the output layer, and RELU activation functions. We considered a batch size of $16$. 

In \cite{Zou2020-ta}, the training data consisted of 100,000 samples, yet we used only 30,000 samples for our simulations and achieved comparable prediction accuracy (see Results).
We compare the prediction accuracy of our neural network models described in Section \ref{sec:NN_model} (LSTM and OptLSTM) with four phylogenetic inference methods: three standard phylogenetic methods 1) Neighbor-joining (NJ) under a Dayhoff amino acid model of evolution \cite{ape, saitou1987neighbor, studier1988note}; 2) Maximum likelihood (ML) with RAxML under a Dayhoff amino acid model of evolution \cite{stamatakis2014raxml}; 3) Bayesian inference (BI) with MrBayes under a Dayhoff amino acid model of evolution with flat priors \cite{huelsenbeck2001mrbayes, ronquist2003mrbayes}, and 4) a previous neural network model by Zou \textit{et al} \cite{Zou2020-ta}. The ``LSTM" and ``OptLSTM" in our results differ only in the network structures: for ``LSTM", we feed each descriptor output into an LSTM network, resulting in 3 separate NN models, while for ``OptLSTM", we run all 3 output of descriptors in one LSTM network by reshaping the tensors. The idea for both model is exactly the same except that ``OptLSTM" is more efficient.
Since the development of this work, two new neural network models have been published \cite{Suvorov2020-pl, leuchtenberger_distinguishing_2020}, yet these two models were implemented for DNA sequences, not proteins, so we were unable to compare the performance of our models with these two neural network implementations (see Discussion).

In each experiment, we set the first 9,000 samples as the training data and the last 1,000 samples as the test data. We train the model using all training data from different dataset, thus including all sets of branch lengths in our simulation. We then test our model for each testing set separately.
For our implementation, we train the data for 300 epochs and test every 10 epochs to record the best accuracy and parameters of our model. We use the Adam optimizer with a learning rate of 0.001 to update our model and we decay the learning rate of each parameter group by 0.9 every 10 epochs.   For the Zou NN implementation \cite{Zou2020-ta}, we use the same parameters, except for the loss function: the reduction applied to the output is summed instead of averaged. For the three standard phylogenetic methods (NJ, ML, BI), we test on the last 500 samples.  

We highlight that for the standard phylogenetic methods, we use the correct model specification. That is, we use the Dayhoff model to estimate the evolutionary distances in NJ with \texttt{phangorn} package in R \cite{Schliep-2011}, and we specify the substitution model to be Dayhoff for RAxML and MrBayes. 
This means that the performance of these methods reflects ideal conditions that are likely unattainable with real data for which the real model is unknown.

\subsection{Results on four taxa}

\begin{figure}[!h]
\centering
\includegraphics[scale=0.5]{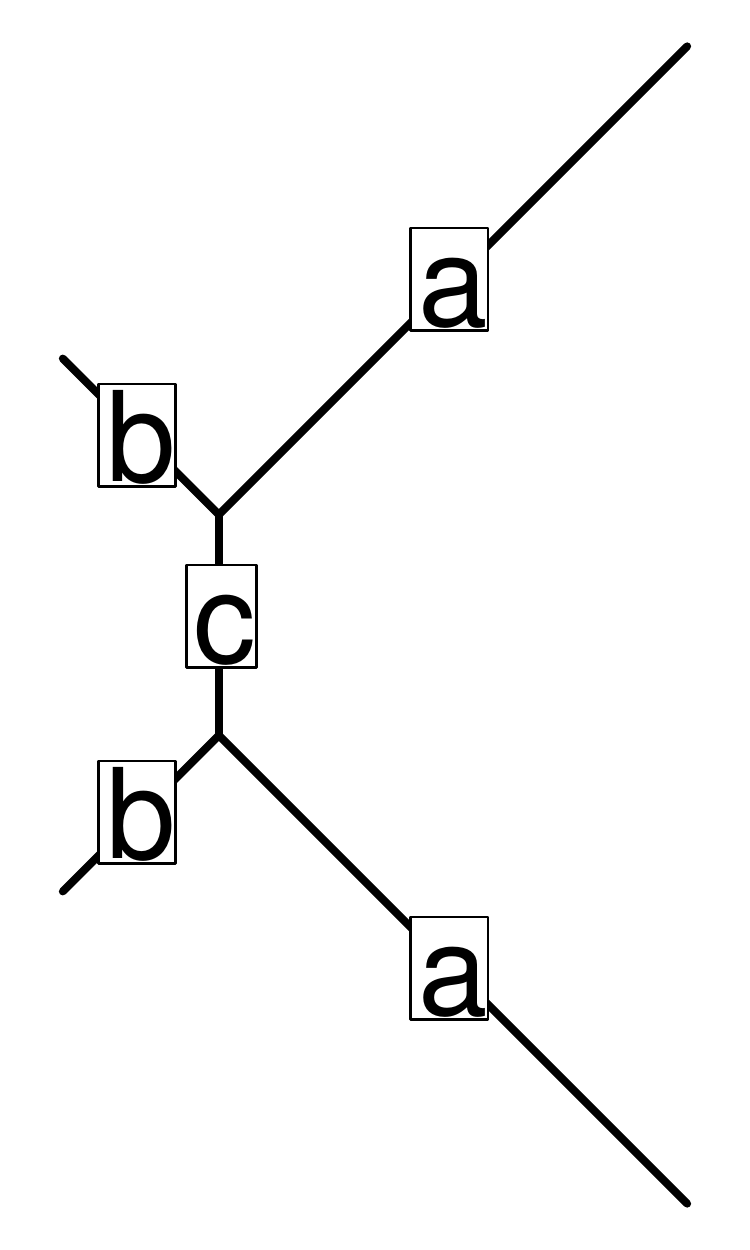}
\includegraphics[scale=0.45]{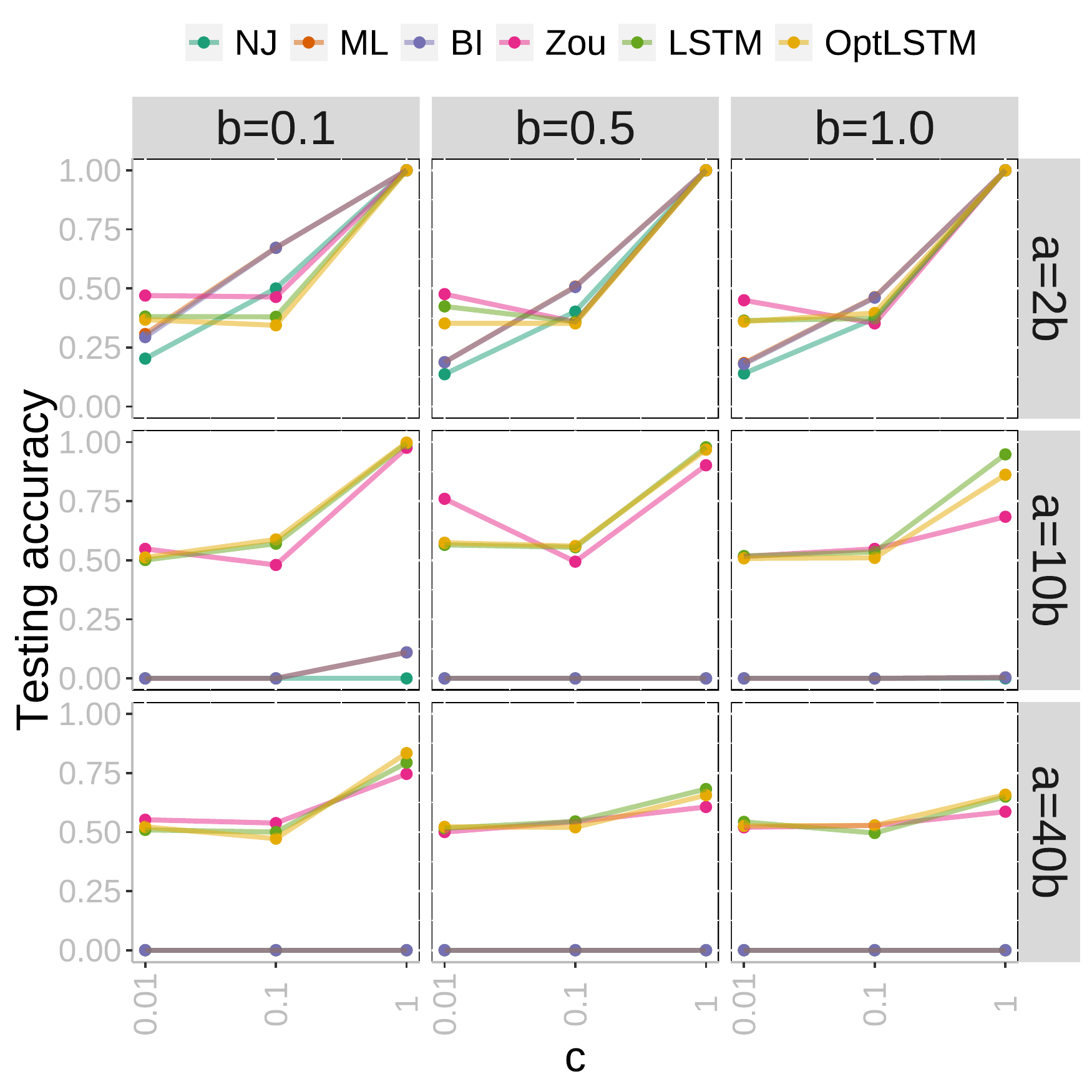}
\caption{Comparison of testing accuracy for six methods: Neighbor-joining (NJ), Maximum Likelihood (ML), Bayesian Inference (BI), Neural Network model in \cite{Zou2020-ta} (Zou), our Neural Network models (LSTM and OptLSTM). We observe that neural network models (Zou, LSTM, OptLSTM) outperform standard phylogenetic inference methods (NJ, ML, BI) on LBA cases ($a=10b, a=40b$). Among those cases, our NN outperforms Zou's with a slightly higher testing accuracy.}
\label{fig:acc}
\end{figure}

Fig.~\ref{fig:acc} (bottom) shows the performance of the six methods to estimate a 4-taxon quartet (top) under a variety of normal and LBA scenarios. We observe that for cases with low LBA ($a=2b, c=1.0$), all six methods perform relatively well with testing accuracy of 100\% in most cases. However, when the branch length $a$ grows ($a=10b,40b$), the standard phylogenetic methods (NJ, ML, BI) are no longer able to estimate the correct quartet tree with prediction accuracy of 0.0 in many of these cases. The neural network implementations, on the other hand, are able to correctly predict 60\% trees even in extreme cases such as $a=40b, c=0.1$, vastly exceeding the naive prediction of 33\%. 

Among the neural network implementations, our model slightly outperforms Zou's \cite{Zou2020-ta}, but while the improvement on prediction accuracy is small, our NN implementation is far more efficient in terms of memory given that we do not create all possible permutations of the input sequences. Using the A100-SXM4 GPU on the platform of Center for High Throughput Computing at University of Wisconsin-Madison, our NN model used on average 11700 megabytes of memory and 3772 megabytes of GPU memory during the training and testing process, which is 18\% and 50\% of the memory and GPU memory used respectively by the Zou's NN model (62670 megabytes of memory and 7415 megabytes of GPU memory on average) with identical training and testing samples and same number of epoches. The efficiency also translated into much faster training with an average of 3.3 hours to train our NN implementation compared to 40.2 hours for the case of Zou's, only 8\% of the time used by Zou's model.

\begin{figure}[!h]
\centering
\includegraphics[scale=0.3]{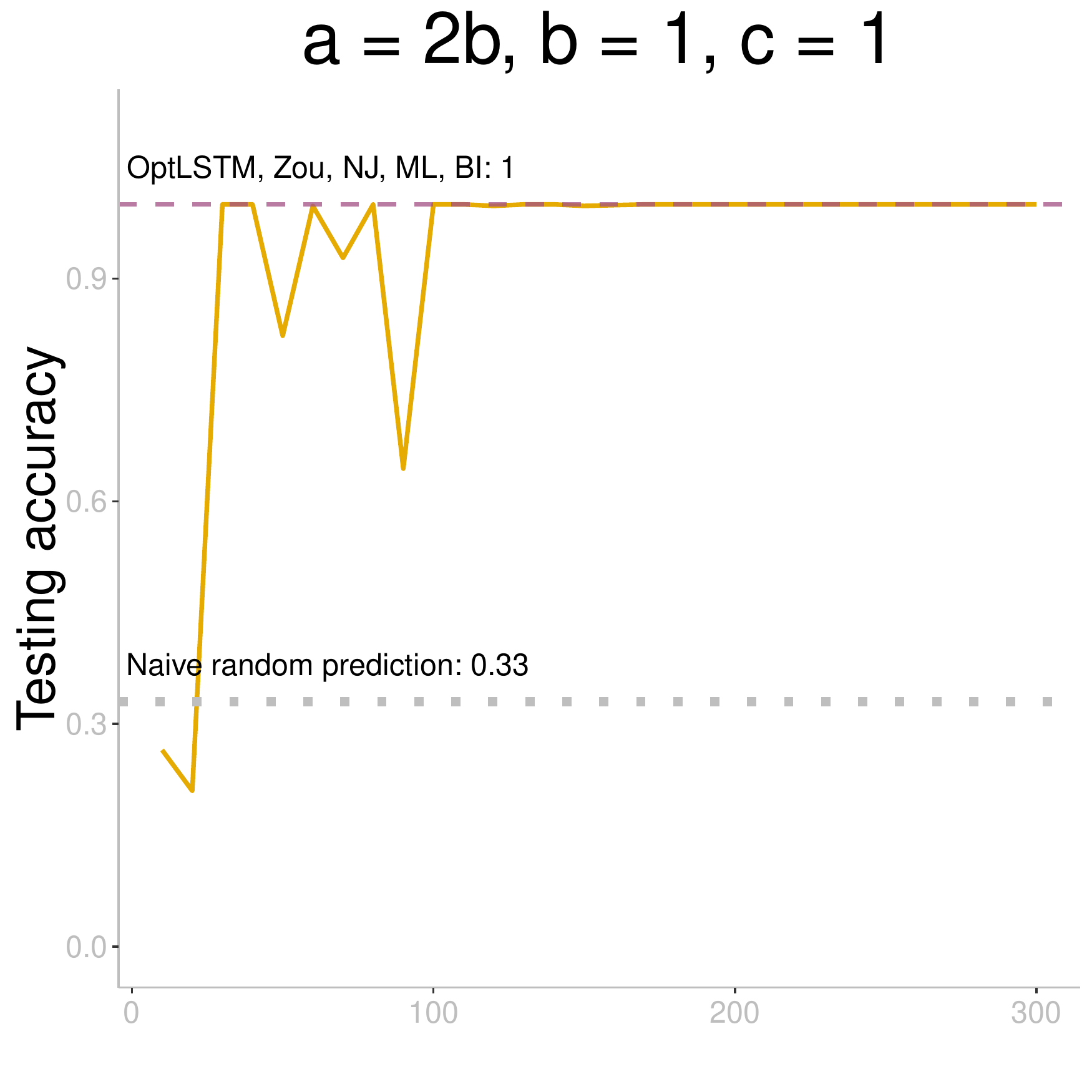}
\includegraphics[scale=0.3]{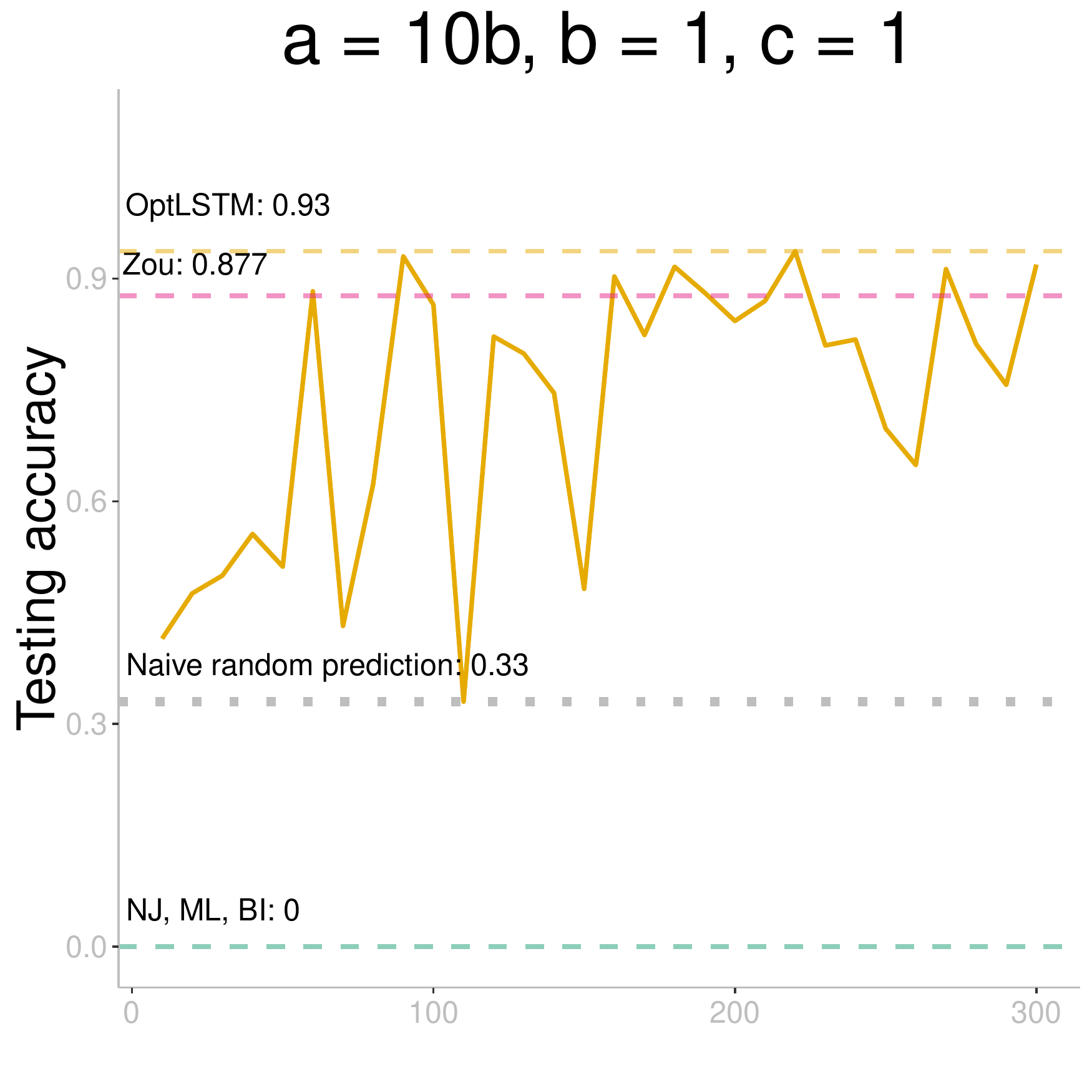}
\includegraphics[scale=0.3]{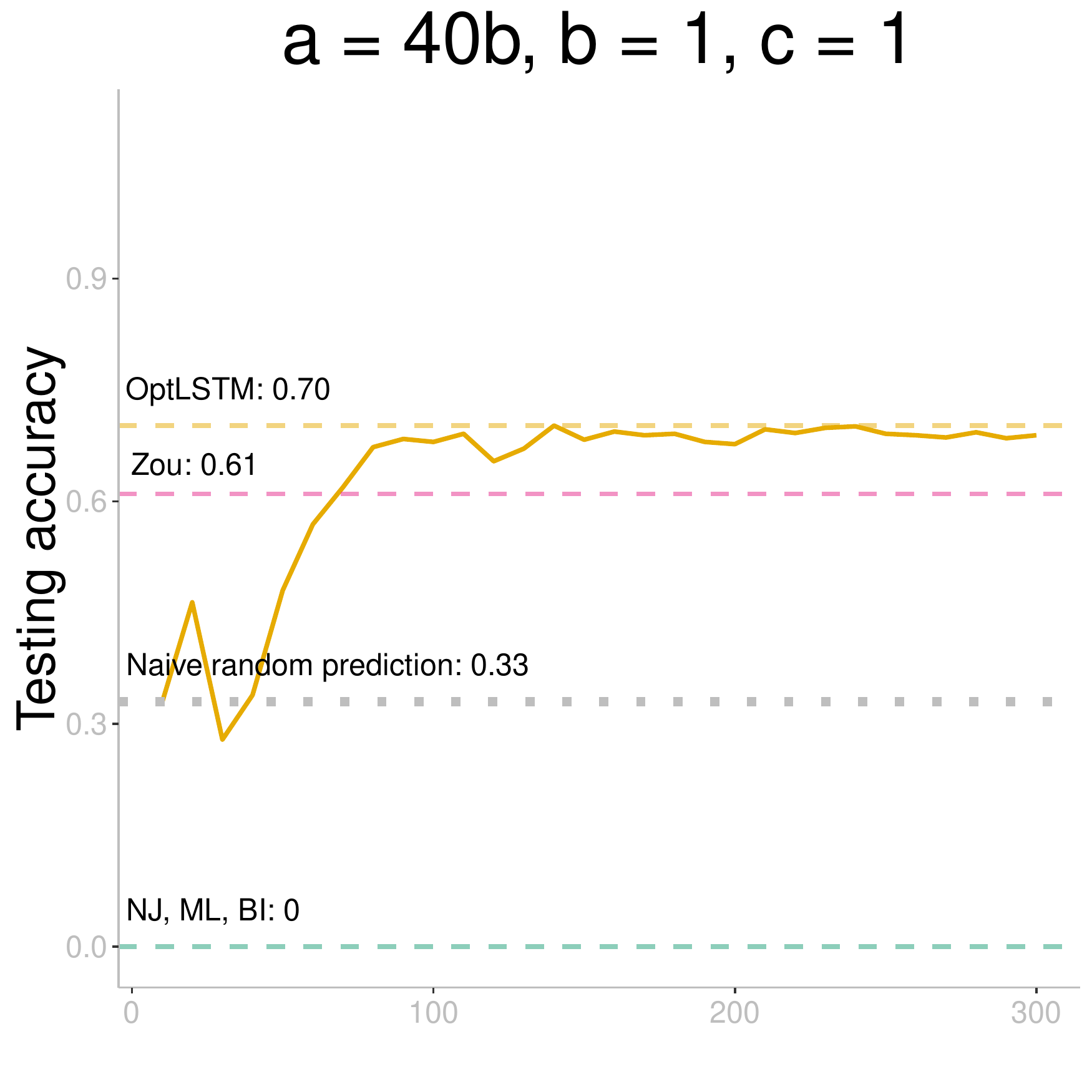}
\caption{Testing accuracy of our NN model implementation (OptLSTM) by epoch for three datasets of increased severity of LBA (top to bottom). Dashed lines represent final prediction accuracy for the other phylogenetic inference methods. For the low LBA case (left: $a=2,b=c=1$), all methods reach 100\% accuracy quite fast while for the other two cases: medium LBA (center: $a=10,b=c=1$) and high LBA (right: $a=40,b=c=1$), all standard phylogenetic methods (NJ, ML, BI) have a 0\% accuracy and the two NN models (Zou in dashed pink line and our OptLSTM model in solid yellow line) have a higher accuracy than the naive random prediction (dotted gray line at 33\%). We note that our method slightly outperforms Zou's NN.}
\label{fig:dynamics}
\end{figure}

Fig.~\ref{fig:dynamics} shows the learning dynamics of our OptLSTM model on three chosen datasets: $c=1, b=1$ with $a=2b,10b,40b$ that represent the three levels of LBA. In these plots, we present as dashed lines the final testing accuracies of the standard phylogenetic methods (NJ, ML, BI) and Zou's NN model. We also present as a gray dotted line the expected accuracy for a naive model that selects quartets at random (33\% for 3 possible quartets). 

For the case of low LBA ($a=2b$), our model reaches a testing accuracy of 100\% rapidly and all methods in this case have a testing accuracy of 100\% as well. As LBA worsens (left to right in Fig.~\ref{fig:dynamics}), the testing accuracy decreases for all methods, but more dramatically for the standard phylogenetic methods (NJ, ML, BI) that have a testing accuracy of 0.0\% (overlapped dashed lines at zero). Only the two NN models (Zou's and our OptLSTM) are able to reach a testing accuracy greater than the naive predictor with our model slightly outperforming Zou's in both cases ($a=10b, 40b$).

\subsection{Simulation strategy on five taxa}
Since LBA is less of a concern on more than four taxa \cite{graybeal1998better}, we use the same 
simulation strategy as with four taxa, but with a different setup for the branch lengths.
We simulate branch lengths by uniformly sampling from $(0,1)$. All other simulation parameters remain the same.

The hyperparameters of the model, however, are 
modified to achieve better performance.
We increase the embedding dimension $m$ from 80 to 160 and the hidden dimension of the LSTM network from 20 to 40, with 6 layers of LSTM instead of 3. We decrease the batch size from 16 to 8. These changes would increase the training time but could offer better model quality overall \cite{keskar2016}.

We use 10,000 samples for simulations to match the training set of quartet trees. Since the model in \cite{Zou2020-ta} could only work in a quartet tree setting, we could not conduct a comparison of performance like we did in the quartet tree case.
While novel deep learning implementations are suitable for more than four taxa \cite{Smith2022, Nesterenko2022},
these approaches are not one-shot learning. That is, these models have non-machine learning steps in the pipeline and are thus not very comparable with our approach.

In this experiment, we take 1,000 samples from the training set to train the model for each epoch. While in theory having more training samples would result in better model quality, in practice, we find no significant difference in model performance between training with the whole set and training with a subset. Training with a subset vastly improves the training time. We test the model performance with 50 samples randomly selected every 10 epochs. The setting for the optimizer and learning rate is the same as the quartet tree cases. We train the model for a total of 1300 epochs. 

\subsection{Results on five taxa}

\begin{figure}[h]
\centering
\includegraphics[scale=0.3]{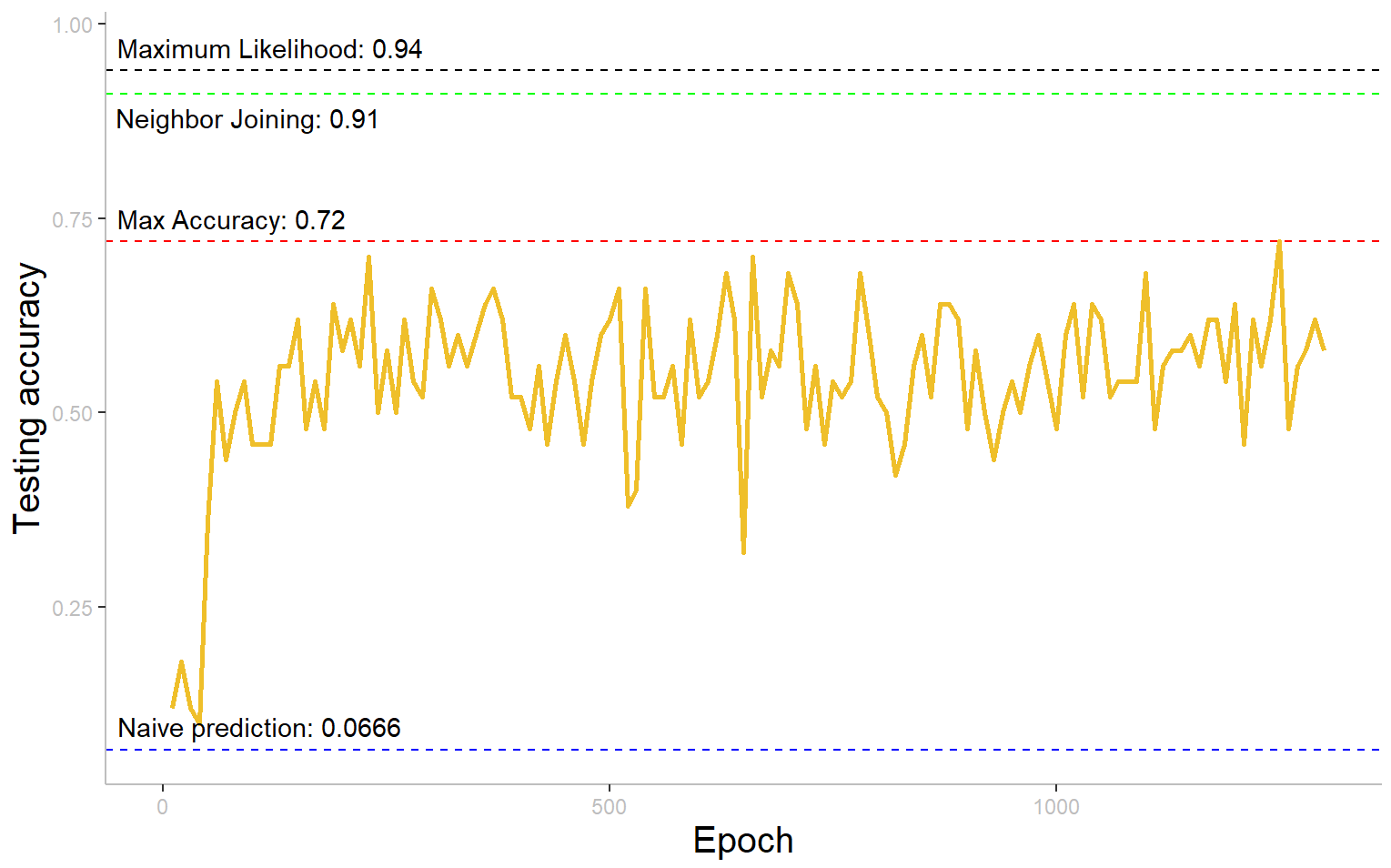}
\caption{Testing accuracy of all 1300 epochs. The black dashed line indicates the result of the Maximum Likelihood method; the green dashed line indicates the result of the Neighbor-Joining method; the red dashed line indicates the maximum accuracy achieved among all epochs; the blue dashed line indicates the random naive prediction. 
}
\label{fig:quintet_eva}
\end{figure}

Fig.~\ref{fig:quintet_eva} shows a 
maximum achieved accuracy of 0.72 with the best model, as shown by the red dashed line. We also show the result from standard phylogenetic methods (ML and NJ) with the black and green dashed lines. It is clear that in this case, the performance of our model is inferior to the standard phylogenetic methods, mostly due to the weakness of the classification task. While the performance is not as good as non-LBA cases for quartet trees, it is worth noting that the classification for a quintet tree is a much harder task than a quartet tree. For a quartet tree, the tree space is 3, thus the naive prediction of the model is 33\%. For a quintet tree, the tree space increases to 15, lowering the naive prediction to 6\%, marked by the blue dashed line in Fig.~\ref{fig:quintet_eva}. Thus, the model has reasonable performance given the number of classes it is trying to classify from. 

\section{Analysis of Zika virus data}
\label{zika}

\begin{figure}[!h]
\centering
\includegraphics[scale=0.6]{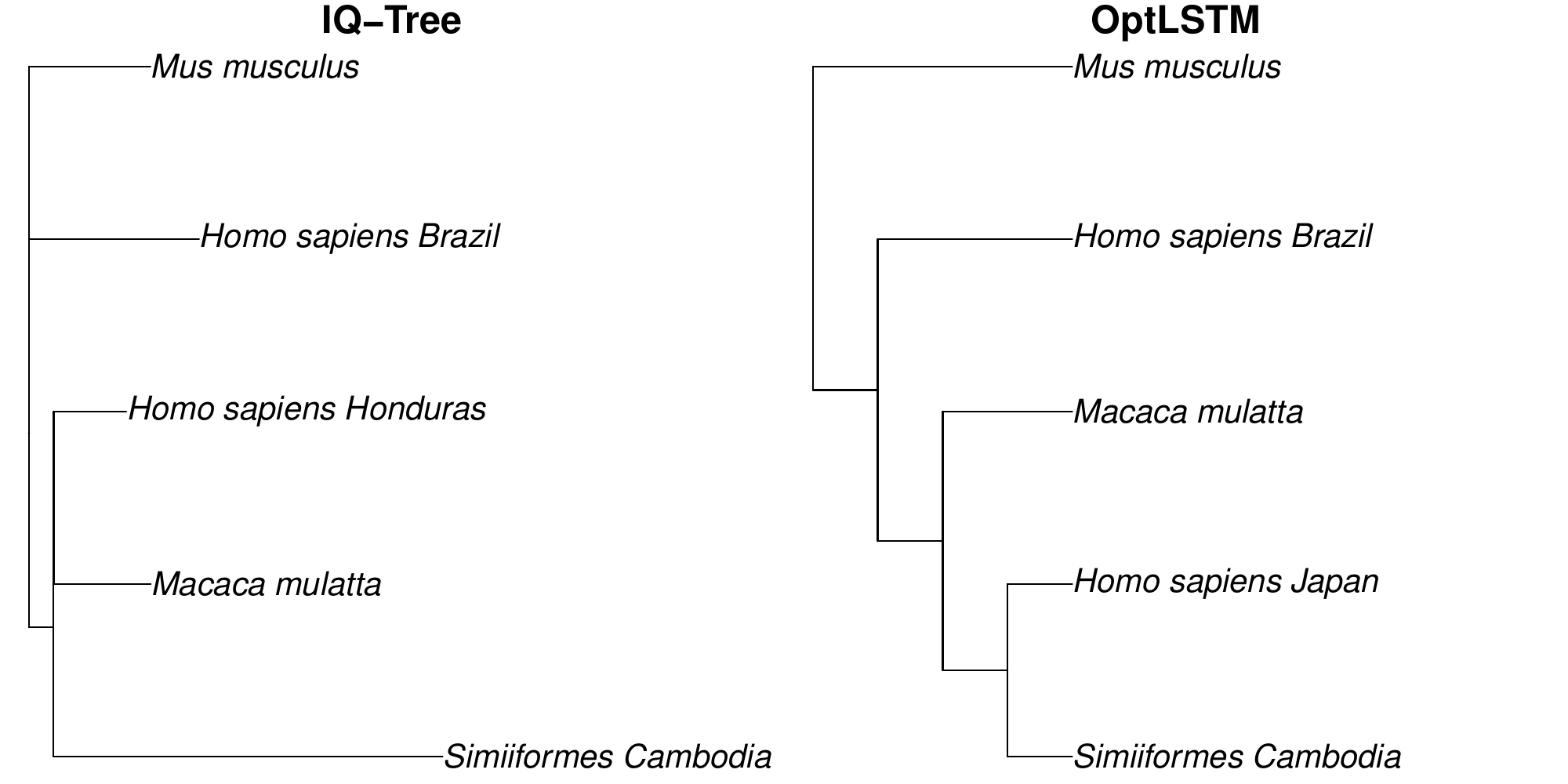}
\caption{Estimated phylogenetic Zika virus tree using IQ-Tree (left) and our OptLSTM model (right). Discrepancies between the trees could be explained by the low phylogenetic signal to resolve the artificial polytomies visible in the IQ-Tree tree. Our tree does not have meaningful branch lenghts as we only infer the tree topology.}
\label{fig:zika}
\end{figure}

Five viral aminoacid sequences for Zika virus were downloaded using NCBI \cite{ncbi} (query link in GitHub repository \url{https://github.com/crsl4/nn-phylogenetics}). See Table \ref{tab:zika} for more details on the sequences. We estimate the best 5-taxon tree with our OptLSTM NN model.
For this case, we train our NN model on a heterogeneous dataset that includes all different scenarios of $a,b,c$ described in the simulations (total sample size 10000). The heterogeneous dataset is constructed by 1000 samples from all different scenarios of $a,b,c$ dataset, and has 10000 samples in total. 
We compare our tree with the estimated tree from IQ-Tree \cite{nguyen2015iq}. We use the default model selection option within IQ-Tree which selected the FLU model \cite{dang2010flu} as the best fit for the data. This model, however, is different from the model we used to train the NN model (Dayhoff). We find that the estimated trees are quite similar regardless of the use of a different model (Figure \ref{fig:zika}). We also find that the IQ-Tree estimated tree has some very short branches which could represent artificial polytomies due to lack of phylogenetic signal to resolve evolutionary relationships.

\renewcommand{\arraystretch}{1.5}
\begin{table}[h]
    \centering
    \caption{Zika viral aminoacid sequences downloaded from NCBI \cite{ncbi}.}
    \begin{tabular}{c|c|c|c|c}
    Accession&Length&Host&Country&Year\\
        \hline
 QIH53581& 3423& \textit{Homo sapiens}& Brazil& 2017\\
 ANG09399& 3423& \textit{Homo sapiens}& Honduras& 2016\\
 AXF50052& 3423& \textit{Mus Musculus}& Colombia& 2016\\
 AWW21402& 3423& \textit{Simiiformes}& Cambodia& 2016\\
 AYI50274& 3423& \textit{Macaca mulatta}& & 2015
    \end{tabular}
    \label{tab:zika}
\end{table}

\section{Discussion}
\label{disc}

We present here the first neural network model that fully exploits the symmetries of trees to avoid permutation operations as other neural models do in phylogenetics \cite{Zou2020-ta}. We showed that our method outperforms standard phylogenetic inference methods like distance-based (NJ), likelihood (ML) and Bayesian (BI) in cases related to LBA, yet it lags behind in the case of five taxa.

While it is difficult to justify the existence of a model that underperforms compared to other phylogenetic inference methods, we argue that the novelty in the mathematical definition of our model is still worthy of recognition. Even when the accuracy of our model to infer 5-taxon trees is below standard phylogenetic methods such as maximum likelihood or neighbor joining, our model is still accurate despite the complexity of the classification task. Furthermore, our model is more efficient in terms of memory and computing time compared to other deep learning implementations. Last, our model is the only existing mathematical representation of a neural network model that intrinsically exploits the tree symmetries, property that can open the door to more complex -- yet efficient -- models in the future.

For example, the main limitation of existing neural network classifiers in phylogenetics \cite{leuchtenberger_distinguishing_2020, Suvorov2020-pl, Zou2020-ta} is that inference is possible only for 4-taxon datasets. Unlike other neural network models, however, our model manages to extend the inference task to 5-taxon trees given the built-in invariance to tree permutations. Zou \textit{et al} \cite{Zou2020-ta} account for these tree isomorphisms by explicitly adding more samples with all the possible permutations of the sequences ($4!=24$ permutations). This data extension imposes demanding memory requirements which become unsustainable for more taxa in the data. Our model, on the contrary, accounts for the tree isomorphisms internally via invariant descriptors which is memory efficient as the number of 
taxa increases. 

Here, we also shed light into the infeasibility of neural network classifiers for larger trees.
The limitation of all existing NN classifiers is that it is very hard to go beyond 5-taxon trees, as our model is an optimized complete search of the tree space. For 6-taxon trees, the tree space is 105, while for bigger trees such as 10-taxon trees, the tree space is 2,027,025. It quickly becomes infeasible in terms of computation time and resources for deep Learning models to search the whole tree space. Most current deep learning models use quartet amalgamation methods such as in Zou \textit{et al}'s model \cite{Zou2020-ta} that combines quartet trees to estimate larger trees. Zaharias \textit{et al} \cite{Zaharias2022-pl} showed that such two stage approach performs worse than direct estimation methods such as ML and NJ. The limitation of deep Learning methods showed in Zaharias \textit{et al}'s work calls for models that avoid doing a complete search on the tree space. While other approaches such as \texttt{Phyloformer} \cite{Nesterenko2022} could estimate larger trees, it is not a direct estimation of trees from sequence data using only the deep learning model. Recently Smith \textit{et al} \cite{Smith2022} proposed a General Adversarial Network (GAN) that does a heuristic search approach of tree space without visiting all possible trees. This approach is very similar to standard phylogenetic inference methods and has achieved good results for up to 10 taxon trees. Future developments of deep learning models for phylogenetic inference should follow this direction by doing heuristic search on the tree space or conditional generation from a learned tree distribution under an unsupervised setting. Another direction would be deep learning models that are designed to work with non-Euclidean space, namely the Graph Neural Network (GNN) \cite{zhang2023learnable, pmlr-v80-jin18a, Jiang2021, Kwon2019}. Such models would more effectively learn the tree topologies of phylogenetic datasets and could potentially offer better performance compared to existing classifiers, which mostly transform phylogenetic trees into Euclidean space datasets and feed into a conventional Deep Learning model such as CNN and RNN.

In addition to extending the model to more taxa, future work will also involve allowing for nucleotide sequences as input in addition to proteins. Currently, our model is only suited to handle protein sequences so that we could compare our performance to \cite{Zou2020-ta}. The other two neural network models applied to phylogenetics \cite{leuchtenberger_distinguishing_2020, Suvorov2020-pl} use nucleotide sequences as input, so we were unable to compare them to our model.
Thus, one future extension of our model could be the modification of the model to allow for the set of nucleotides to be used as the dictionary for the sequences. Furthermore, Suvorov \textit{et al} \cite{Suvorov2020-pl} allow for gaps in the alignment and proved that a neural network model is better suited to handle highly gapped sequences (which are quite common on bacterial or viral sequences due to difficulty of alignment) compared to standard phylogenetic methods. Therefore, another extension of our model will be the incorporation of gaps in the sequences to test its performance to a variety of gap scenarios.

\section*{Data Availability}
All the scripts for the model implementation, simulations and real data analysis can be found in the GitHub repository: 
\url{https://github.com/crsl4/nn-phylogenetics}.

\section*{Acknowledgements}

This work was partially supported by the Department of Energy [DE-SC0021016 to C.~S.-L.], National Science Foundation [DEB-2144367 to CSL] and the National Science Foundation [DMS-2012292 to L.~Z.-N.]. 

\bibliography{references}  

\end{document}